\documentclass[compsoc]{IEEEtran}
\usepackage{graphicx}
\usepackage{balance}  
\usepackage{amsmath}
\usepackage{pxfonts}

\title{Full Speed Ahead: 3D Spatial Database Acceleration with GPUs}

\begin{document}
\title{Full Speed Ahead: 3D Spatial Database Acceleration with GPUs}

\author{Lucas C. Villa Real\footnote{Corresponding author: lucasvr@br.ibm.com}, Bruno Silva\\\{lucasvr, sbruno\}@br.ibm.com\\IBM Research}

\date{8 June 2018}
\maketitle

\begin{abstract}
Many industries rely on visual insights to support decision-making processes in
their businesses. In mining, the analysis of drills and geological shapes,
represented as 3D geometries, is an important tool to assist geologists on the
search for new ore deposits.  Aeronautics manipulate high-resolution geometries
when designing a new aircraft aided by the numerical simulation of
aerodynamics. In common, these industries require scalable databases that
compute spatial relationships and measurements so that decision making can be
conducted without lags. However, as we show in this study, most database
systems either lack support for handling 3D geometries or show poor performance
when given a sheer volume of data to work with.  This paper presents a
pluggable acceleration engine for spatial database systems that can improve the
performance of spatial operations by more than 3000x through GPU offloading. We
focus on the design and evaluation of our plug-in for the PostgreSQL database
system.
\end{abstract}

\section{Introduction}
Database systems are a fundamental component of software architectures that
deal with the storage and retrieval of digital information. They are used
on a variety of scenarios that include the management of large-scale datasets,
data curation, and analysis. Data can be represented as a collection of tables
that are manipulated through relational operators (as in the traditional
relational database systems) or as unstructured documents that are accessed
without a rigid schema, as seen in key-value or graph-based NoSQL databases.

As database systems offer generic interfaces to manage a large set of
data types, it is expected that custom support for special classes of data
may be missing, incomplete, or poorly implemented. One example of such a
class is geospatial data; with a few exceptions, database systems are not
readily able to distinguish geometries from other binary blobs nor to provide
operators to enable filtering based on spatial properties and relationships
between geometry types.

The geospatial services industry is estimated to have an economic impact
at USD 1.6 trillion anually in revenues. The technology enables companies
to sharpen decision making and save time and money~\cite{geospatial2016whereisthemoney}.
Given the importance that geospatial data plays, several databases provide
extensions to enable their base systems to handle spatial data types.
That is the case of MySQL (with MySQL Spatial Extension, supporting 2D
geometries), IBM's DB2 (with its Spatial Extender for 2D data),
PostgreSQL (that enables operations on 2D and 3D geometries through the
PostGIS extension), and Oracle Database (that does the same with
its Oracle Spatial extension)~\cite{piorkowski2011spatialstandards}.
In common, such extensions enable the representation of geometries
as if they were native data types (i.e., integer, floating point, text) and
include a data model, new indexing methods, predicate functions, and topology
operations, among others~\cite{stolze2003sqlmm}.

Despite the presence of spatial data types on modern databases, processing
of 3D spatial queries often suffer from poor scalability on such platforms.
The issue arises from a combination of an intensive demand for computational resources
with the presence of very large collections of 3D objects. As a consequence,
individual query evaluations can take several hours to complete or -- worse --
end up timing out and giving no results back to the user~\cite{zhang2009spatialmapreduce}.
Such limitations severely restrict the applications of 3D spatial analysis
supported by traditional database systems.

We present in this paper the design and evaluation of a database-agnostic
accelerator for 3D spatial queries that removes the scalability problems
aforementioned. Our prototype features specialized GPU kernels that perform
spatial operations on geometries comprising millions of polygons and on
a virtually unlimited number of objects. Our evaluation of the platform,
based on a comparison with the open source PostGIS spatial extension to
PostgreSQL, shows performance gains of more than 3000$\times$ on several use
cases. We also describe a connector for PostgreSQL that enables the routing
of queries over PostGIS functions and objects to our platform.

The paper is organized as follows. Section~\ref{sec:databases} gives an
overview of database extensions that handle spatial data types and that
enable access to foreign services. Next, Section~\ref{sec:gpu_accel}
presents the core architecture of our accelerator. Section~\ref{sec:performance}
compares the performance of our platform against CPU-based implementations.
We present related work in Section~\ref{sec:relatedwork} and make closing
remarks thereafter in Section~\ref{sec:conclusions}.

\section{Database Extensions}
\label{sec:databases}
The work described in this paper is based on two particular standards that
have been incorporated into PostgreSQL. This section describes their purpose
and how they relate to us.

\subsection{Foreign Data Wrappers}

A part of the SQL standard known as SQL/MED offers syntax extensions to SQL
that enable access to data that lives outside the database through the table
interface. There are no restrictions on where the foreign data must be stored:
it can live on another database, on a remote filesystem, inside a structured
file, or in any other repository determined by the user -- as long as the remote
server can handle the SQL/MED protocol and map the data to a tabular layout.

The SQL/MED interfaces allow a SQL server to receive a query that contains
mixed references to local and remote tables. In that case, the SQL server
decomposes the query into multiple fragments and dispatches them to the foreign
servers as needed. It also devises and initiates an execution plan for each
fragment, receives the responses from each remote server, and returns the
result to the application~\cite{melton2002sqlmed}.

Our acceleration platform implements a subset of the SQL/\-MED protocol that
enables its integration with a regular database system. The data it exposes
to the SQL server mirrors the original tables containing geometry data types,
with a few restrictions. First, the mirror is populated on demand from the
original SQL server. Second, the mirrored data is kept in memory in a format
that can be readily parsed by the GPU kernels. Third, the only members from
the original table that are mirrored are (i) the geometry data and (ii) its
unique identifier; all operations involving other table members such as
numerical or textual ones are processed by the SQL server itself. These
decisions allowed us to focus on the spatial aspects of our engine with
no need to implement a full featured database management system.

\subsection{Spatial Data Types}

There exist two major standards supporting the inclusion of spatial data
types and operators into SQL databases. The first, SQL/MM (ISO/IEC 13249),
extends SQL to address simple 2D elements and 3D geometries. It supports
the management of collections of points, curves, surfaces, and a combination
of primitives and spatial methods involving geometry sets~\cite{stolze2003sqlmm}.
The second standard is OGC Simple Feature Access, defined by the Open
Geospatial Consortium. It supports the storage, retrieval, query and updates
of feature collections in 2 or 3 spatial dimensions~\cite{chen2008spatialextensions}.

Our accelerator prototype supports a subset of the OGC standard: it handles
geometries represented by points, triangulated irregular networks (TINs),
polyhedral surfaces, and line strings. Three spatial operators are currently
implemented: 3D Intersection, 3D Distance, and 3D Volume. Together, they
cover the majority of the use cases whose performance we intend to improve.


\section{Accelerating Spatial Databases}
\label{sec:gpu_accel}
Here we describe the architecture of our platform, including its coupling
with a SQL database system, and the algorithms we implemented to enable
the efficient processing of selected spatial operations on GPUs.

\subsection{Architecture}

The acceleration platform implements a subset of the PostgreSQL protocol
that handles user authentication, the execution of simple queries and
prepared statements (including support for JOIN expressions and nested
selections), and the retrieval of results through named cursors and
anonymous portals. It can be accessed via regular PostgreSQL
tools or through foreign data facilities.

\begin{figure}[!t]
        \centering
        \includegraphics[width=.98\linewidth]{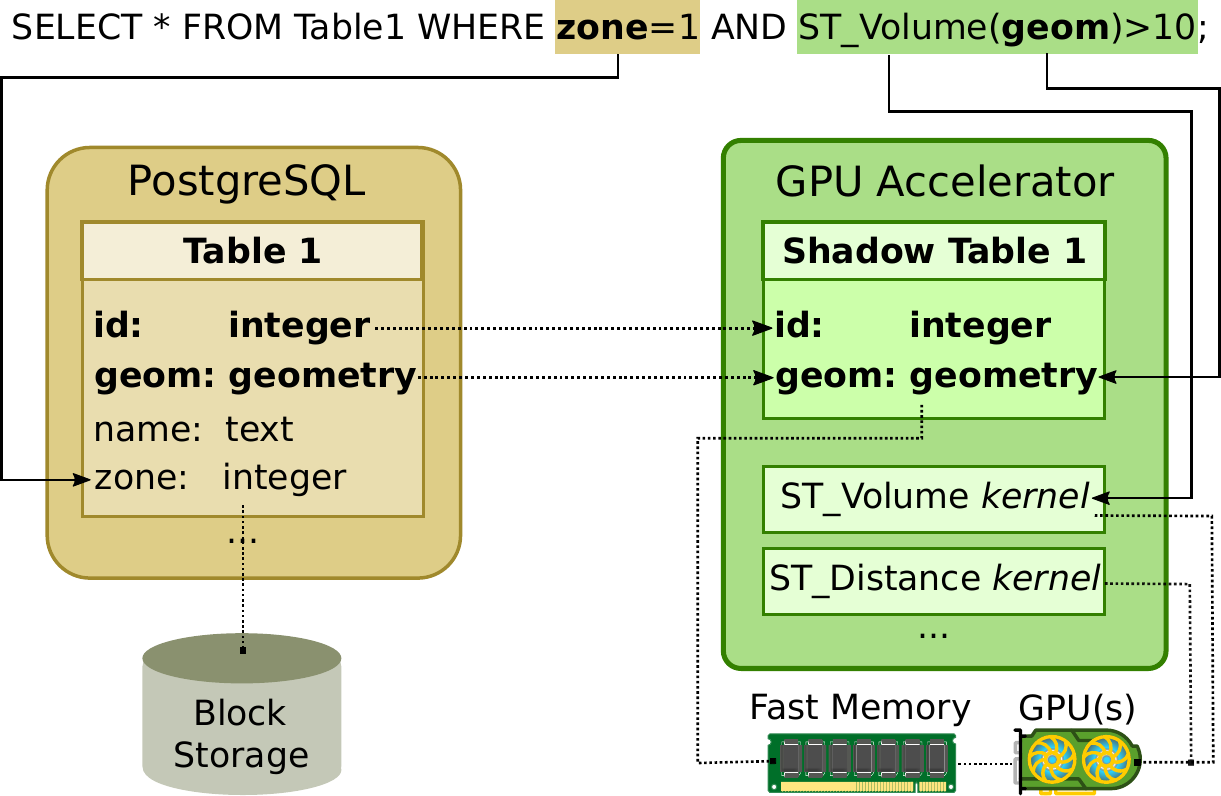}
        \caption{%
            Queries are automatically split based on the presence of spatial
            operators. Regular operations are handled by the traditional
            PostgreSQL pipeline, whereas the spatial components are dispatched
            to our acceleration engine. The associated geometry data and
            identifiers are cached in memory and processed by specialized
            GPU kernels.
        }
        \label{fig:arch}
\end{figure}

An overview of our architecture is given in Figure~\ref{fig:arch}. Two
major components are shown: an instance of PostgreSQL, holding a table
with a geometry and its attributes, and an instance of the GPU accelerator.
The latter mirrors the geometry and its unique identifier from PostgreSQL;
other table members are not carried over.
Queries submitted to the PostgreSQL server are split according to the presence
of foreign elements. On the depicted example, the non-spatial element of the query is
handled directed by the PostgreSQL server, whereas the ST\_Volume operation
on the geometry column executes on the GPU accelerator.

The accelerator platform retrieves the geometry data from the original table
and persists that data in system memory. This process is conducted asynchronously
either on demand (as queries arrive) or at startup time. All data featured on the
geometry column is processed by the GPUs, even in the presence of ``WHERE'' clauses
that could restrict the computation to a smaller geometry set. We do so to prevent
sub-optimal use of cores within the GPU's streaming multiprocessors and to cache
results of computations that may be asked in the near future. Under this platform,
SQL ``WHERE'' clauses, if given, execute on the CPU over the GPU kernel's
output. The results are consolidated by the PostgreSQL server once both sides
have finished executing their jobs.

\subsection{GPU Algorithms}

In this section, we present how the volume, distance, and intersection algorithms are implemented on the GPU.

\subsubsection{Volume}

\begin{figure}[!t]
        \centering
        \includegraphics[width=.98\linewidth]{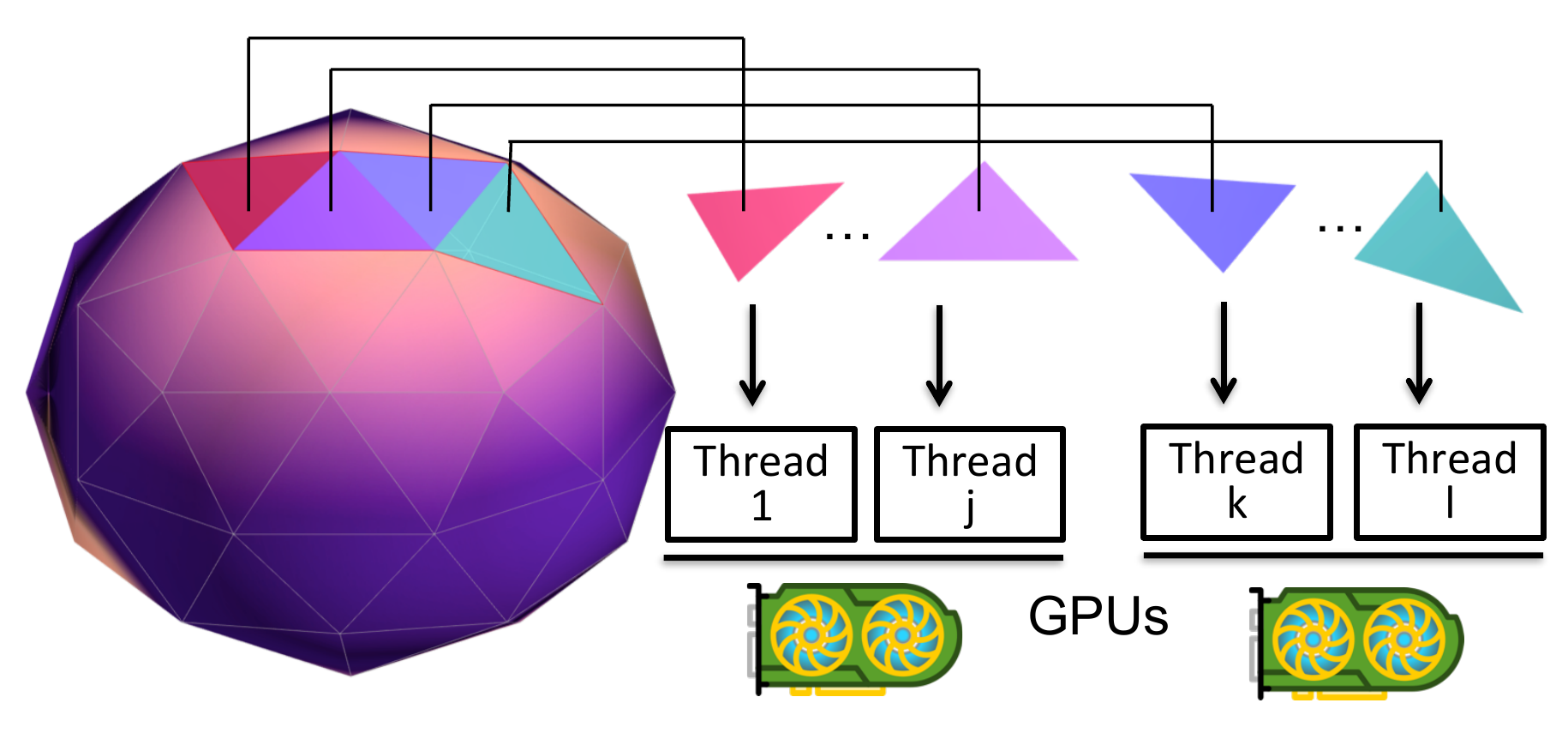}
        \caption{%
            Example on how to evaluate the volume of a 3D triangle meshes using
            GPUs.  According to the divergence theorem, it is possible to
            calculate the volume of a given solid mesh by using the normal of
            each face. Each GPU thread computes the area and the normal of each
            mesh triangle and these results are combined to compute the mesh
            volume. Other operators like distance and intersection use a
            similar approach in the sense that each GPU thread performs some
            basic computation on a solid face.
        }\label{fig:acceleration_idea}
\end{figure}

The proposed GPU accelerator employs the divergence theorem
\cite{gold2018spatial} to evaluate the volume of a 3D mesh indirectly by using
the information related to each polygon face. The theorem is defined as 

\begin{equation}
    \oiiint_V {(\nabla \cdot \mathbf{F})} dV = \oiint_{S}{(\mathbf{F} \cdot \mathbf{n})} dS,
\end{equation}

where $V$ represents a volume in $\mathbb{R}^3$, $S$ is the boundary of V ($S =
\partial V$), $\mathbf{F}$ is a continuously differentiable vector field (flux) defined
on a neighborhood of $V$, and $\mathbf{n}$ is the unit normal pointing outward $S$. 

Let $P$ be a polyhedron representing a database geometry given by a set of
triangles $A_i$, $i = 0, ..., N - 1$ with vertices $(u_i, v_i, w_i)$. As all
the faces are counter clockwise on $A_i$, the outer unit normal $n$ to $P$ on
each face $A_i$ is $n_i = \hat{n}_i \div |\hat{n}_i|$ where $\hat{n} = (v_i -
u_i) \times (w_i - u_i)$. Therefore, if we assume a flux $F = \vec{p}/3$ where
$\vec{p}$ is a point on the surface of P, then the volume ($V$) of $P$ can be
evaluated as 

\begin{equation}
\begin{aligned}
    V &= \int_P{1} dP = \int_P{(\nabla \cdot \vec{p}/3)} dP =  \frac{1}{3} \int_{\partial P} (\vec{p} \cdot n) d{(\partial P)} \\
    &= \frac{1}{3} \sum_{i=0}^{N-1}\int_{A_i} (u_i \cdot n_i) d{A_i} = \frac{1}{6} \sum_{i=0}^{N-1} u_i \cdot \hat{n}_i,
\end{aligned}
\end{equation}

where we exploit the fact that the area of $A_i$ is $|\hat{n}|/2$, and $\vec{p}
\cdot n_i$ is constant over each $A_i$. Figure~\ref{fig:acceleration_idea}
presents the strategy to accelerate the volume calculation of 3D polyhedral
meshes. Differently from a sequential CPU execution, GPU threads calculate $u_i
\cdot \hat{n}_i$ for each face $A_i$ in parallel. Depending on the number of
faces of the solid, its volume can be computed in a single GPU parallel
execution. The implementations of distance and intersection operators follow a
similar approach as each GPU thread performs a basic computation on the solid's
face.

\subsubsection{Distance}

The current implementation of 3D distance supports the following geometries:
(i) line segment/line segment, (ii) line segment/polyhedral surface, and (iii)
point/polyhedral surface. The following explanation is related to the distance between line
segment and polyhedral surface; the other distance variations employ an
analogous approach. In order to find the distance between a line segment and a
polyhedral surface we find the distance between each triangular face of the polyhedron and the line segment. 
Similarly to the volume calculation, the distance of each
face and the line is calculated in a parallel GPU thread. Then, the minimum
distance is returned to the user.

We employ the approach suggested by \cite{eberly2007, schneider2002geometric}
to evaluate the distance between a triangle and a line segment. Assume a line
segment with end points $P_0$ and $P_1$ with parametric representation $L(t) =
P_0 + t\vec{d}$, $0 \leq t \leq 1$ where $\vec{d} = P_0 - P_1$. Let a triangle
with vertices $V_0, V_1, V_2$ be represented as $T(u, v) = V_0 + u\vec{e_0} +
v\vec{e_1}$ where $\vec{e_0} = V_1 - V_0$, $\vec{e_1} = V_2 - V_0$, $0 \leq u,
v \leq 1$, and $u + v \leq 1$. The minimum distance is computed by locating the
values $u, v, t$ so that $T(u, v)$ is the triangle point closest to $L(t)$. We
find $u, v, t$ that minimizes the squared-distance $Q(u, v, t)$ (Equation
\ref{eq:distanceTL}) from $T$ to $L$ to find the minimum distance from $L$ to each mesh
triangle $A_i$.

\begin{equation}
\label{eq:distanceTL}
Q(u, v, t) = |T(u, v) - L(t)|^2
\end{equation}
    
\subsubsection{Intersection}

The intersection supports the same geometries and utilizes the same face
decomposition approach as the 3D distance. Whenever polyhedral meshes are
involved in the computation, the operator decomposes the intersection
evaluation of each polyhedral face in a GPU thread. We employ a less
computationally-intensive evaluation for intersection when compared to the
distance operator. 

For instance, for the line segment and polyhedral surface intersection, each
GPU thread intersects the line segment with the plane containing the given
triangular face, and then determines whether or not the intersection point is
within the triangle~\cite{eberly2007}. In this case, we
represent the triangle with vertices $V_0$, $V_1$, and $V_2$ as $T(u, v, w) =
uV_0 + vV_1 + wV_2$ where $w = 1 - u - v$, $0 \leq u, v \leq 1$, $u + v \leq
1$. The triple $(u, v, w)$ is known as barycentric coordinates of $T$
\cite{schneider2002geometric}. Assume also a line segment with points $P_0$ and
$P_1$ and parametric representation $L(t) = P_0 + t\vec{d}$, $0 \leq t \leq 1$
where $\vec{d} = P_0 - P_1$. Then the values for $t, u,$ and $v$ should be  

\begin{equation}
  \begin{bmatrix}
    t \\ 
    u \\
    v \\
  \end{bmatrix}
    = \frac{1}{(\vec{d} \times e_1)\cdot(e_0)} 
  \begin{bmatrix}
      ((P - V_0) \times e_0) \cdot e_1 \\ 
      (\vec{d} \times e_1) \cdot (P - V_0) \\ 
      ((P - V_0) \times e_0) \cdot \vec{d} \\ 
  \end{bmatrix},
\end{equation}

where $e_0 = V_1 - V_0$ and $e_1 = V_2 - V_0$. If $t$, $u$, and $v$ respect the aforementioned contraints, then $L$  intersects $T$.

\section{Performance Evaluation}
\label{sec:performance}
\textbf{Use case:} the query acceleration platform has been evaluated on a synthetic
data set that represent features from the mining domain~\cite{our_dataset}.
The data contains three object types: (i) line segments 
representing drill holes, (ii) closed meshes representing the locations and areas of
ore bodies, and (iii) block models used for mineral resource estimation. We restrict
our tests to a single ore shape comprising 500 faces and to 5 million drill
holes. The queries we run represent spatial operations that geologists perform on a
daily basis: computing the volume of geological shapes, filtering drill holes based
on their distance to profitable areas of a mine, and retrieving drill holes that
intersect with certain objects of interest (such as a particular rock type).

\textbf{Computing environment:} our experiments were run on a bare metal machine with
an Intel E5-2620 v4 featuring 16 cores, 256 GB of memory, 800 GB of SSD, and a NVIDIA
Tesla V100 GPU. The software stack includes PostgreSQL 10.4 and PostGIS 2.4.4
with the SFCGAL backend (version 1.3.2) to enable the ST\_Volume operator.
Our GPU kernels were built with the Cuda compiler version 9.1.85. The PostgreSQL
server and our accelerator were hosted on that same machine. Kernel page caches
are flushed prior to each execution.

\textbf{PostgreSQL settings:} we changed the default amount of memory dedicated for
caching data from 80 MB to 50 GB, enforced the use of
parallel processing, configured the cost of non-sequentially-fetched disk pages to
1, and modified the estimated cost of the PostGIS functions to enable their execution
with a varying number of workers.

\begin{figure}[!t]
        \vspace{-2mm}
        \centering
        \includegraphics[width=1\linewidth]{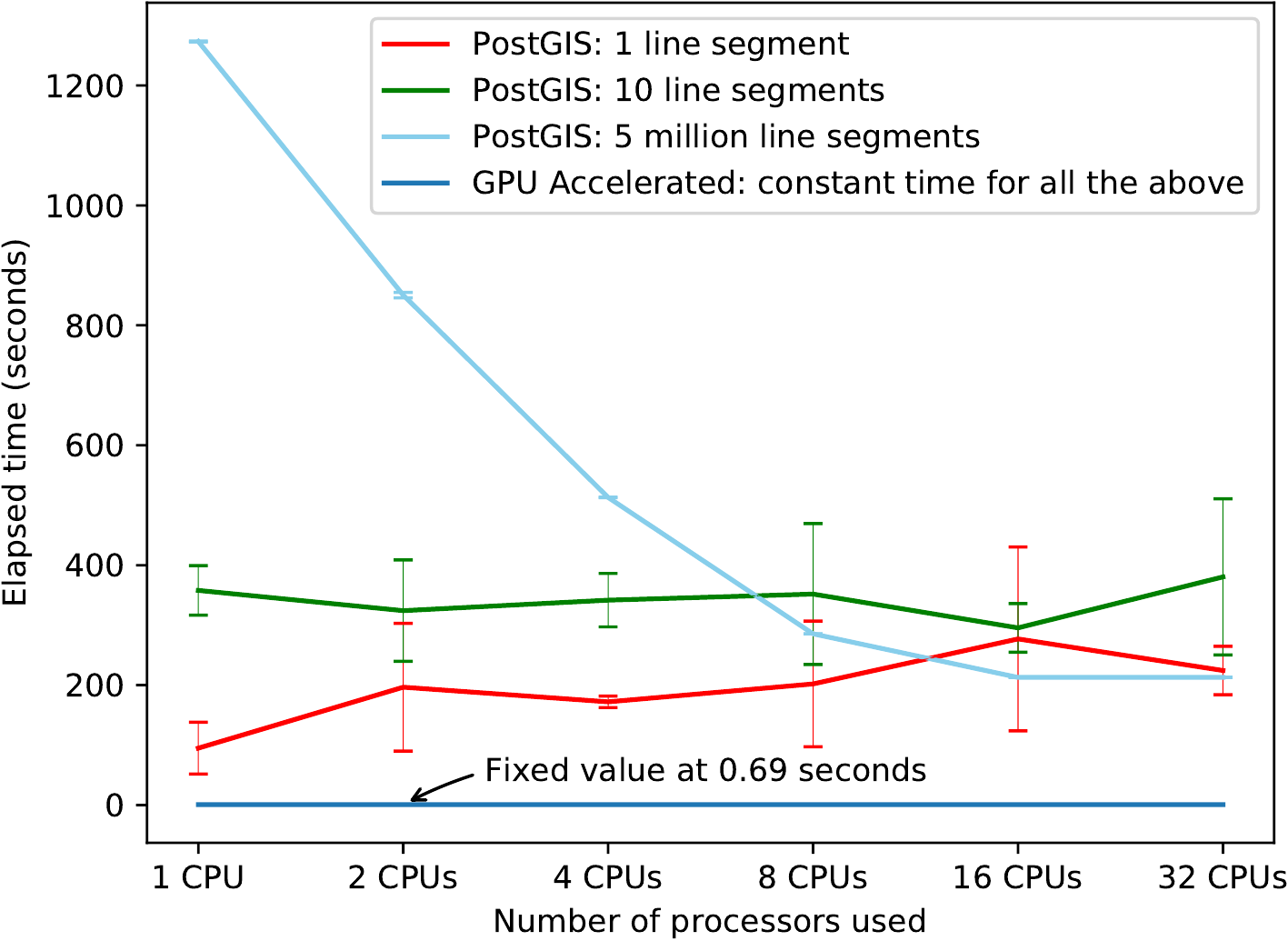}
        \caption{%
        Variable performance of 3D distance computation on the CPU versus a
        constant time with our GPU accelerator.
        }\label{fig:perf_distance}
        \vspace{-2mm}
\end{figure}

\textbf{Results:} we separate the results in three sets: volume, distance,
and intersection. Figure~\ref{fig:perf_distance} shows the processing time to
determine the distance of line segments to a 3D solid. PostGIS
shows a significant variation, especially when few geometries are processed, due to
cache hits in PostgreSQL buffers. We also note similar processing times
with 16 and 32 CPUs. That is caused by PostgreSQL' parallel planner, which picked
11 workers regardless of having enough geometries to keep all cores busy. Performance
gains with pure CPU parallelization were of 6$\times$ when processing the full dataset.
Our GPU-based accelerator, on the other hand, shows a consistent
performance independently of how many geometries are asked by the user (1, 10, or
5 million): 0.685 seconds with a variation of 0.002 seconds. That is an 
\textbf{improvement of 1860$\times$} over the sequential version of PostGIS.

\begin{figure}[!t]
        \vspace{-2mm}
        \centering
        \includegraphics[width=.98\linewidth]{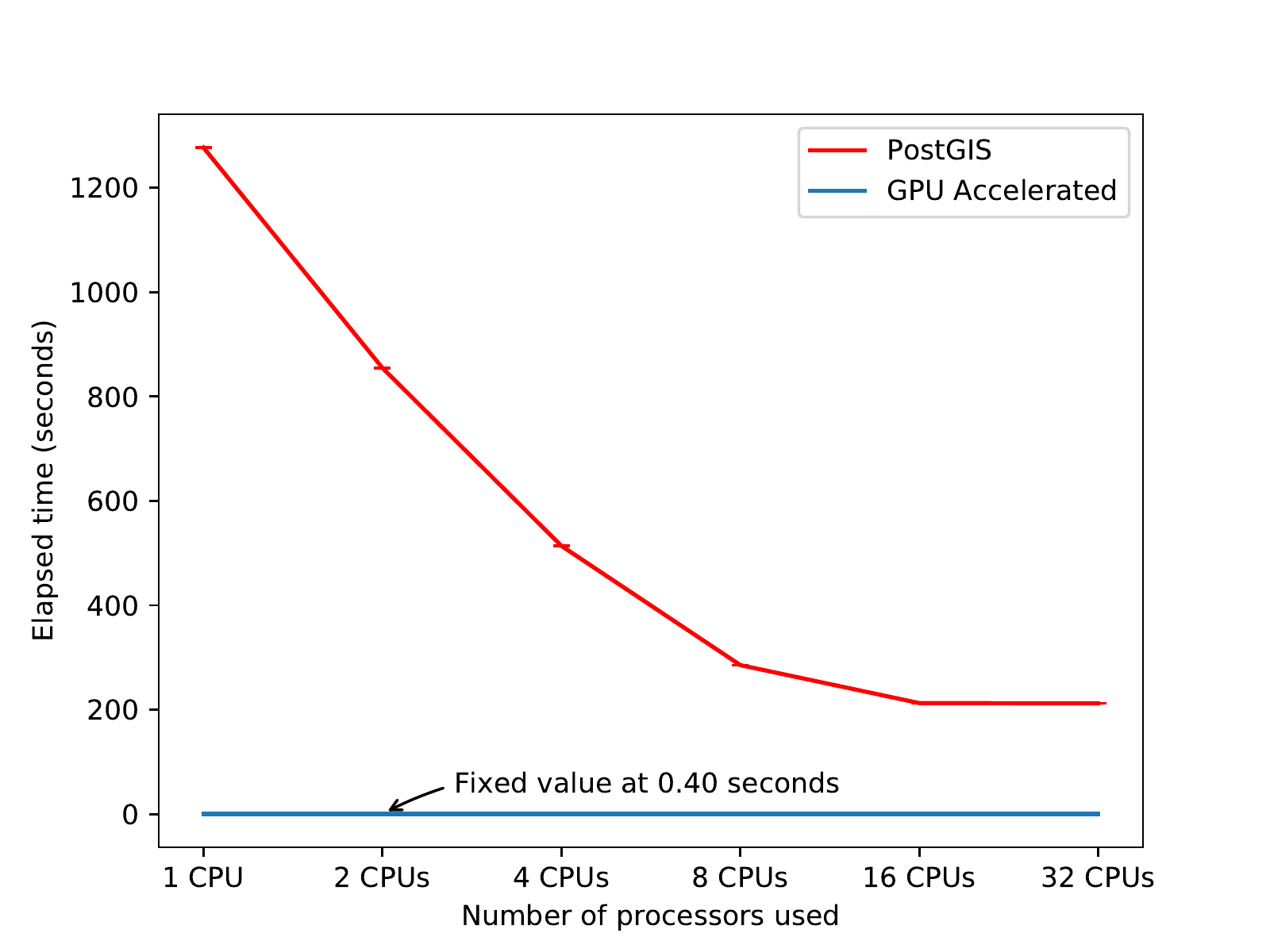}
        \caption{%
        Performance of computing the 3D distance of 5 million line segments and
        a solid on the CPU and the GPU.
        }\label{fig:perf_intersection}
        \vspace{-2mm}
\end{figure}

For the intersection test, shown in Figure~\ref{fig:perf_intersection}, we considered
the 5 million segments case alone. The performance notes above regarding
PostGIS apply again. This time, however, our GPU-based accelerator
\textbf{improves 3230$\times$} over PostGIS' sequential run. As observed in Section
3.2.3, we employ a less computationally intensive algorithm for 3D intersection
when compared to 3D distance, which leads to the dramatic speedup observed here.

We also computed the 3D volume of the same solid used on the other tests. PostGIS
does not split the geometry among different workers, so a single processor is used
at all times. PostGIS computes the volume in 2530 seconds, with a variation of 68 
seconds. Since our GPU algorithm processes different faces on dedicated GPU processors,
we observed a \textbf{gain of 2770$\times$} at 0.91 seconds with a variation of 0.006.

\section{Related Work}
\label{sec:relatedwork}
Several areas of computation (e.g., artificial intelligence)  are leveraging the vast amount of processing power and memory bandwidth provided by modern GPUs for data-intensive applications.

Recently, database researchers also employed GPUs for accelerating data management systems. In \cite{Breb2014}, the authors present a survey for GPU-accelerated database systems and argue that modern database management systems should be an in-memory, column-oriented using a block-at-a-time processing model, possibly extended by a just-in-time-com\-pi\-la\-tion component. These systems also should have a query optimizer that is aware of co-processors (e.g. GPUs) and data-locality. An important problem in query optimization is join-order optimization. Current join-order optimizers employ dynamic programming in a sequential approach. In \cite{meister2016challenges}, the authors propose different ways to employ GPU-accelerated dynamic programming for join-order optimization, and discuss the challenges of this approach. Sitaridi et. al. propose an implementation of relational operators on GPU processors. Their focus is related to string matching in SQL queries. As GPU threads in the presence of different execution paths are serialized, they split string matching into multiple steps to reduce thread divergence. Their solution optimizes string search by selecting a given parallelism granularity and string layout for different algorithms.

\section{Conclusions}
\label{sec:conclusions}
The work presented in this paper shows that it is possible to obtain expressive
performance gains from spatial queries (by more than 3000$\times$) by coupling
existing database systems with a standalone accelerator supported by GPUs. As
we conducted this work we identified several research opportunities, including
geometry caching strategies, GPU-assisted data compression, pre-fetching
algorithms, and cooperation between concurrent GPU kernels. We intend to
investigate each of these topics in the near future.

\balance

\bibliographystyle{IEEEtran}
\bibliography{references}

\begin{thebibliography}{10}
\providecommand{\url}[1]{#1}
\csname url@samestyle\endcsname
\providecommand{\newblock}{\relax}
\providecommand{\bibinfo}[2]{#2}
\providecommand{\BIBentrySTDinterwordspacing}{\spaceskip=0pt\relax}
\providecommand{\BIBentryALTinterwordstretchfactor}{4}
\providecommand{\BIBentryALTinterwordspacing}{\spaceskip=\fontdimen2\font plus
\BIBentryALTinterwordstretchfactor\fontdimen3\font minus
  \fontdimen4\font\relax}
\providecommand{\BIBforeignlanguage}[2]{{%
\expandafter\ifx\csname l@#1\endcsname\relax
\typeout{** WARNING: IEEEtran.bst: No hyphenation pattern has been}%
\typeout{** loaded for the language `#1'. Using the pattern for}%
\typeout{** the default language instead.}%
\else
\language=\csname l@#1\endcsname
\fi
#2}}
\providecommand{\BIBdecl}{\relax}
\BIBdecl

\bibitem{geospatial2016whereisthemoney}
A.~Datta, ``\emph{Where is the money in geospatial industry?}'' 2016,
  https://www.geospatialworld.net/article/where-is-the-money (visited on Jun 6,
  2018).

\bibitem{piorkowski2011spatialstandards}
A.~Pi\'orkowski, ``{MySQL Spatial and PostGIS - Implementations of Spatial Data
  Standards},'' in \emph{Electronic journal of Polish agricultural
  universities}, vol.~14, 01 2011, p.~03.

\bibitem{stolze2003sqlmm}
K.~Stolze, ``{SQL/MM Spatial: The Standard to Manage Spatial Data in Relational
  Database Systems},'' in \emph{Proceeding of the 10th Conference on Database
  Systems for Business, Technology, and Web ({BTW})}, 2003, pp. 247--264.

\bibitem{zhang2009spatialmapreduce}
S.~Zhang, J.~Han, Z.~Liu, K.~Wang, and S.~Feng, ``{Spatial Queries Evaluation
  with MapReduce},'' in \emph{2009 Eighth International Conference on Grid and
  Cooperative Computing}, Aug 2009, pp. 287--292.

\bibitem{melton2002sqlmed}
\BIBentryALTinterwordspacing
J.~Melton, J.~E. Michels, V.~Josifovski, K.~Kulkarni, and P.~Schwarz,
  ``{SQL/MED: A Status Report},'' \emph{SIGMOD Rec.}, vol.~31, no.~3, pp.
  81--89, Sep. 2002. [Online]. Available:
  \url{http://doi.acm.org/10.1145/601858.601877}
\BIBentrySTDinterwordspacing

\bibitem{chen2008spatialextensions}
\BIBentryALTinterwordspacing
R.~Chen and J.~Xie, \emph{Open Source Databases and Their Spatial
  Extensions}.\hskip 1em plus 0.5em minus 0.4em\relax Berlin, Heidelberg:
  Springer Berlin Heidelberg, 2008, pp. 105--129. [Online]. Available:
  \url{https://doi.org/10.1007/978-3-540-74831-1\_6}
\BIBentrySTDinterwordspacing

\bibitem{gold2018spatial}
C.~Gold, \emph{Spatial Context: An Introduction to Fundamental Computer
  Algorithms for Spatial Analysis}, ser. ISPRS Book Series.\hskip 1em plus
  0.5em minus 0.4em\relax CRC Press, 2018.

\bibitem{eberly2007}
D.~Eberly, \emph{3D Game Engine Design: A Practical Approach to Real-Time
  Computer Graphics}, ser. Morgan Kaufmann series in interactive 3D
  technology.\hskip 1em plus 0.5em minus 0.4em\relax Taylor \& Francis, 2007.

\bibitem{schneider2002geometric}
P.~Schneider and D.~Eberly, \emph{Geometric Tools for Computer Graphics}, ser.
  The Morgan Kaufmann Series in Computer Graphics.\hskip 1em plus 0.5em minus
  0.4em\relax Elsevier Science, 2002.

\bibitem{our_dataset}
K.~Maestrini, ``\emph{Synthetic dataset representing mine features},''
  http://lucasvr.gobolinux.org/publications/2018-ADMS-Dataset.zip (visited on
  Jul 22, 2018) ~.

\bibitem{Breb2014}
\BIBentryALTinterwordspacing
S.~Bre{\ss}, M.~Heimel, N.~Siegmund, L.~Bellatreche, and G.~Saake,
  \emph{GPU-Accelerated Database Systems: Survey and Open Challenges}.\hskip
  1em plus 0.5em minus 0.4em\relax Berlin, Heidelberg: Springer Berlin
  Heidelberg, 2014, pp. 1--35. [Online]. Available:
  \url{https://doi.org/10.1007/978-3-662-45761-0\_1}
\BIBentrySTDinterwordspacing

\bibitem{meister2016challenges}
A.~Meister and G.~Saake, ``{Challenges for a GPU-Accelerated Dynamic
  Programming Approach for Join-Order Optimization.}'' in \emph{In Proc.
  GI-Workshop GvDB}, 2016, pp. 81--86.

\end{thebibliography}

\end{document}